\documentclass[prl,twocolumn,floatfix,showpacs,tightenlines]{revtex4}
\usepackage{psfig}
\begin{document}

\title{Optical sum rules that relate to the potential energy of strongly 
correlated systems}
\author{J.~K.~Freericks$^*$, T.~P.~Devereaux$^\dagger$, M. Moraghebi$^\dagger$,
 and S.~L.~Cooper$^\ddagger$}
\affiliation{$^*$Department of Physics, Georgetown University, Washington, DC
20057, U.S.A.}
\affiliation{$^\dagger$Department of Physics, University of Waterloo, Canada}
\affiliation{$^\ddagger$Department of Physics, University of Illinois,
Urbana-Champaign, IL}

\date{\today}

\begin{abstract}
A class of sum rules for inelastic light scattering is developed. We show that 
the first moment of the non-resonant response provides information about 
the potential energy in strongly
correlated systems. The polarization dependence of the sum rules provide
information about the electronic excitations in different regions of the 
Brillouin zone. We determine the sum rule for
the Falicov-Kimball model, which possesses a metal-insulator
transition, and compare our results to the light scattering experiments in
SmB$_6$.
\end{abstract}


\maketitle
One of the key signatures of strong electron correlations in
condensed matter systems is the redistribution of spectral weight
from low to high energies.  A classic example of this phenomenon 
occurs when a normal metal becomes a superconductor at low temperatures.
The optical conductivity is
suppressed at low energy due to the presence of the
superconducting gap~\cite{richards_tinkham}.  Because there is an
optical sum rule---the $f$-sum rule---which constrains the integrated spectral 
weight in $\sigma(\nu)$, the spectral weight suppressed below the 
superconducting energy gap reappears as a delta function peak at 
$\omega=0$, reflecting the onset of many-body coherence in the system.
In high temperature superconductors, low-energy spectral weight has also 
been observed to shift to high energies below the 
superconducting transition~\cite{htsc_oc}. This dramatic spectral weight 
transfer is believed to be a signature of the strong electronic correlations 
in these and related materials~\cite{Osafune}.

The use of sum rules in optical conductivity measurements has had a wide impact in a
number of fields of science.  
The optical sum rule originated in atomic physics as a relationship between the 
total number of electrons in the atomic system and the integrated spectral
weight~\cite{f-sum-rule}. 
In a solid state system, the optical sum rule is usually projected onto 
the lowest energy band.  In this case, the integrated spectral weight 
associated with electrons in the lowest band is related to the average 
kinetic energy of the electrons, regardless of the nature of the 
electronic interactions~\cite{maldague}.  Since the electronic kinetic
energy usually varies on an energy scale on the order of electron
volts, the average kinetic energy is essentially a constant at low
to moderate temperatures, and so the optical conductivity sum rule is 
essentially temperature-independent.

It would be useful to have similar sum rules that relate to the potential
energy.  In strongly correlated electron systems in which electronic 
interaction energies are on the order of, or greater than, the kinetic 
energy, efforts to study the potential energy as functions of doping, 
pressure, or temperature have been hindered by the limited tools available 
for measuring dynamical properties.
We show here that the non-resonant inelastic light scattering response has sum 
rules similar to the f-sum rule that relate to the potential
energy, and that these sum rules should provide 
valuable insight into the 
scattering of optical or x-ray photons from materials~\cite{RIXS}.
As an added benefit compared to optical conductivity measurements, 
inelastic light scattering allows one to probe the 
electronic excitations in different regions of the Brillouin 
zone (BZ) by orienting the incident and scattered light 
polarizations.

Shastry and Shraiman~\cite{shastry_shraiman} were the first to
point out that, if the self energy is local and $d\rightarrow\infty$, the 
optical conductivity and the 
nonresonant Raman response function in the depolarized ($\epsilon_I$ 
perpendicular to $\epsilon_S$) scattering geometry are related according to 
$\sigma(\nu) \propto S(\nu)/\nu$ a relationship that was proved in 
Ref.~\cite{freericks_devereaux}. This result
implies that there is also a sum rule for Raman scattering, which is 
related to the f-sum rule for optical conductivity and to the average 
kinetic energy of the electrons.

In this paper, we show that there are also interaction-dependent sum rules 
for inelastic light scattering that are related to the potential energy for 
${\bf q}=0$ Raman scattering, and to combinations of the potential and 
kinetic energies for ${\bf q}\ne 0$ inelastic x-ray scattering.

The formalism used to derive these sum rules is
straightforward. If we consider the time-ordered product that
yields the susceptibility corresponding to the fluctuations of an
operator ${\mathcal O}$
\begin{equation}
\chi(\tau)={\rm Tr}\langle {\mathcal T}_\tau e^{-\beta {\mathcal H}}
{\mathcal O}(\tau){\mathcal O}^\dagger(0)\rangle/{\mathcal Z}-\left (
{\rm Tr}\langle e^{-\beta {\mathcal H}}{\mathcal O}\rangle/\mathcal{Z}
\right )^2,
\end{equation}
where $\tau$ is the imaginary time, ${\mathcal H}$ is the Hamiltonian,
$\beta$ is the inverse temperature, ${\mathcal Z}={\rm Tr}
e^{-\beta{\mathcal H}}$ is the partition function
and ${\mathcal O}(\tau)=e^{\tau{\mathcal H}}{\mathcal O}
e^{-\tau{\mathcal H}}$. Then it is easy to show from the spectral formula
that when the susceptibility is Fourier transformed, and analytically
continued to the real frequency axis, we get
\begin{equation}
\int_0^{\infty}d\nu\nu{\rm Im}\chi(q,\nu)=\frac{\pi}{2}{\rm Tr}\langle e^{-\beta
{\mathcal H}}[{\mathcal O},[{\mathcal H},{\mathcal O}^\dagger]]\rangle/
{\mathcal Z}
\label{Sum_Rule}
\end{equation}
as the general sum rule for the first moment of the susceptibility
(the brackets indicate commutators). If we consider the polarization
operator ${\mathcal O}_{\rm pol}=e\sum_{i}{\bf R_{i}}n_{i}$ with $n_{i}$ the 
number density
operator at site $i$, then this gives the $f$-sum rule where
$\sigma(\nu)={\rm Im}\chi_{JJ}(\nu)/\hbar\nu$ in terms of the current-current
correlation function $\chi_{JJ}$ (with $J=i[{\mathcal H},{\mathcal O}_{\rm pol}]/\hbar$:
\begin{equation}
\int_{0}^{\infty}d\nu\sigma(\nu)=\frac{\pi e^{2}a^{2}}{2\hbar}{\rm Tr}
\langle e^{-\beta {\mathcal H}}(-\hat T)\rangle/{\mathcal Z}.
\end{equation}
The sum rule is proportional to the average kinetic energy
(we consider only nearest neighbor hopping on a hypercubic lattice
$\hat T = -t\sum_{i,\delta,\sigma}c_{i,\sigma}^{\dagger}c_{i+\delta,\sigma}$
with $\delta$ a nearest-neighbor translation vector).
Likewise, if we consider the dynamical charge susceptibility
${\mathcal O}_{\rm ch}({\bf q})=
\sum_{k}c^{\dagger}_{{\bf k}+{\bf k}/2,\sigma}c_{{\bf k}-{\bf q}/2,\sigma}$
we obtain
\begin{eqnarray}
&&\int_{0}^{\infty}d\nu \nu {\rm Im} \chi_{ch}({\bf q},\nu)=\nonumber\\
&&\frac{\pi}{2}
Tr\langle e^{-\beta H}\sum_{\bf k}[\epsilon({\bf k+q})+\epsilon({\bf k-q})
-2\epsilon({\bf k})]\rangle /{\mathcal Z},
\nonumber\\
&&=\pi (1-X){\rm Tr}\langle e^{-\beta{\mathcal H}}(-\hat T)
\rangle/{\mathcal Z},\quad {\rm (in~}d\rightarrow\infty{\rm )}
\end{eqnarray}
with $X=\lim_{d\rightarrow\infty}\sum_{i=1}^d\cos(k_{i})/d$ and 
$\epsilon({\bf q})$ the 
band structure. This vanishes as $q^{2}$ for ${\bf q}$=0 as it must,
due to total charge conservation. 

There is, however, no charge conservation for inelastic light scattering
with ${\bf q}=0$ optical photons (Raman scattering). Raman
scattering can be classified into representations of the irreducible
point group symmetry of the crystal, selected by orienting the incident and 
scattered polarization
vectors. The light scattering polarizations allow energy fluctuations to be 
projected onto different
regions of the BZ, which has been useful in elucidating the behavior of
electron dynamics for different momentum states~\cite{old_papers}. For example, 
for nonresonant scattering in tetragonal systems, the operator 
$\mathcal{O}_{A_{\rm 1g}}({\bf q})= \sum_{\bf k} \epsilon({\bf k})
c^\dagger_{{\bf k}+{\bf q}/2} c_{{\bf k}-{\bf q}/2}=\hat T({\bf q})$  is the 
kinetic energy operator for ${\bf q}=0$ $A_{\rm 1g}$ scattering, and 
${\mathcal O}_{B_{\rm 1g}}({\bf q})= \sum_{\bf k} \epsilon({\bf k}+{\bf Q})
c^\dagger_{{\bf k}+{\bf q}/2}c_{{\bf k}-{\bf q}/2}$ is a modified 
kinetic energy operator for ${\bf q}=0$ $B_{\rm 1g}$ scattering for the zone boundary 
wavevector ${\bf Q}=(\pi,0)$ [which is 
generalized to higher dimensions as ${\bf Q}=(\pi,0,\pi,0,...)$]. 
Thus the $A_{1g}$ scattering response is associated with energy fluctuations 
from all regions of the BZ, while the $B_{\rm 1g}$ response is associated with 
dynamics near the BZ edges and away from the diagonals.  The sum
rule in Eq.~(\ref{Sum_Rule}) for nonresonant light scattering
is similar to the optical sum rule, except the
(model-dependent) operator average is different, and now depends
on the potential energy since ${\mathcal H}=T+V$ [see Eq.~(\ref{Sum_Rule})]. 
In the general case, there are two contributions---one from 
the kinetic energy commutator and one from the potential energy commutator:
\begin{equation}
\int_{0}^{\infty}d\nu \nu {\rm Im} \chi_{in}({\bf q},\nu)=\frac{\pi}
{2{\mathcal Z}} {\rm Tr}
\langle e^{-\beta{\mathcal H}}[\hat T({\bf q,Q}) +\hat U({\bf q,Q})]\rangle.
\label{sr}
\end{equation}
The first term on the right-hand side of Eq.~(\ref{sr}) does not depend on the 
type of Hamiltonian and is given by
$\hat T({\bf q,Q})=-\sum_{{\bf k},\sigma} t({\bf k,q,Q})c^{\dagger}_{{\bf k},\sigma}
c_{{\bf k},\sigma}$, with
$t({\bf k,q,Q})=\epsilon({\bf k-Q-q/2})^{2}[\epsilon({\bf k-q})-\epsilon({\bf k})]
+\epsilon({\bf k-Q+q/2})^{2}[\epsilon({\bf k+q})-\epsilon({\bf k})]$. Regardless
of the symmetry [${\bf Q}=0$ for $A_{1g}$ or $(\pi,0)$ for $B_{1g}$] the 
kinetic-energy
part of the sum rule does not contribute to the $({\bf q}=0)$ Raman
scattering response, but it does contribute to the $({\bf q}\ne 0)$ 
inelastic X-ray scattering response in a polarization-dependent manner. 

However the second term in Eq.~(\ref{sr}) depends on the particular choice for
the potential energy in the Hamiltonian. 
Evaluating these sum rules for different models is complicated, because the
operator averages involve complex correlation functions.  
We consider the simplest model here,
the spinless Falicov-Kimball model (FK)~\cite{falicov_kimball}
\begin{equation}
H_{FK}=-\frac{t^*}{2\sqrt{d}} \sum_{\langle
i,j\rangle}c^\dagger_ic_j 
+U\sum_ic^\dagger_ic_iw_i .
\label{eq: hamiltonian}
\end{equation}
Here $c^\dagger_i$ ($c_i$) creates (destroys) a conduction electron
at site $i$, $w_i$ is a classical variable (representing the
localized electron number at site $i$) that equals 0 or 1, $t^*$
is a renormalized hopping matrix that is nonzero between nearest
neighbors on a hypercubic lattice in $d$-dimensions~\cite{metzner_vollhardt}, 
$U$ is the local screened Coulomb interaction between 
conduction and localized electrons, and
$\langle i,j\rangle$ denotes a sum over sites $i$ and nearest
neighbors $j$.  In our calculations,
the average filling for conduction and localized electrons
is set to $1/2$.  The FK model can
be thought of as a limit of the Hubbard model for which one species of spin 
is not allowed to hop.  

For the FK model, the operator $\hat U$ in the sum rule is
\begin{eqnarray}
&&\hat U_{FK}({\bf q,Q})=\frac{Ut^{*2}}{2dN}\sum_{{\bf i,\delta,\delta^{\prime}}}c^{\dagger}_{i}c_{i+\delta+\delta^{\prime}}
e^{-i {\bf Q}\cdot({\bf \delta+\delta^{\prime}})}\\
&&\times[e^{-i{\bf\frac{q}{2}}\cdot({\bf \delta+\delta^{\prime}})}(w_{{\bf i+\delta^{\prime}}}-w_{{\bf i+\delta+\delta^{\prime}}})
-e^{i{\bf\frac{q}{2}}\cdot({\bf \delta+\delta^{\prime}})}(w_{\bf i}-w_{{\bf i+\delta}})].\nonumber
\label{FK}
\end{eqnarray}
This operator involves the difference of correlated hopping
operators, where the hopping of the conduction electrons is
correlated with the presence of the localized electrons; note
how the potential energy directly enters when $\delta+\delta^\prime=0$.
The operator for the Hubbard model is similar, but involves complex spin
and spin-flip hopping correlation functions, which could be measured,
in principle, with neutron scattering. It is surprising that the sum
rule for a charge response function can be related to a spin response
function in the Hubbard model.

The result in Eq.~(\ref{FK}) is valid for any dimension. It can be evaluated
exactly by examining the large dimensional limit using results 
from dynamical mean field theory (DMFT)~\cite{fk_rev}. The result for the FK 
model is
\begin{widetext}
\begin{eqnarray}
\int_{0}^{\infty}d\nu \nu &{\rm Im}& \chi_{in}({\bf q},\nu)=\frac{\pi}
{2{\mathcal Z}}{\rm Tr}
\langle \hat T({\bf q,Q}) +\hat U({\bf q,Q})\rangle=\nonumber\\
&& {\rm Im}\int d\omega f(\omega)\left\{ (1-X) X^{\prime 2}
\left[1-\frac{3}{2}Z(\omega)G(\omega)-Z^{2}(\omega)\left[1-Z(\omega)G(\omega)\right]\right]+\frac{1}{2}(X-1)[1-Z(\omega)G(\omega)]\right\}\nonumber\\
&&+{\rm Im}\int d\omega f(\omega)(\Sigma(\omega)-U\langle w_{i}\rangle)
\left\{\frac{1-X^{\prime 2}}{2}G(\omega)+\left\{Z(\omega)-1/G(\omega)\right\}X^{\prime 2}\right\},
\label{eq: final_fk_result}
\end{eqnarray}
\end{widetext}
with $f(\omega)=1/[1+\exp(\beta\omega)]$ the Fermi function, $G(\omega)$ and
$\Sigma(\omega)$ the Green's function and  self energy, respectively, and
$Z(\omega)=\omega+\mu-\Sigma(\omega)$.  The difference between $A_{\rm 1g}$ and
$B_{\rm 1g}$ scattering comes from the momentum factors
$X^{\prime}=\lim_{d\rightarrow\infty}(1/d)\sum_{i=1}^{d}\cos(q_{i}/2)h^{i}$, 
with $h=1$ ($-1)$ for $A_{\rm 1g}$
($B_{\rm 1g}$) scattering. 
Note that both the first term associated with the kinetic energy and the 
second term associated with the potential energy contribute to the sum rule 
for inelastic x-ray scattering (${\bf q}\ne 0$).  On the other hand, only the 
second ``potential energy'' term contributes to the sum rule for Raman 
scattering (${\bf q}=0$).  Therefore, the sum rule's momentum dependence 
contains information regarding the potential and kinetic energy contributions.

Note that the sum rule for $(X=1)$ Raman scattering is proportional to the
correlated part of the self energy, due to the factor 
$\Sigma(\omega)-U\langle w_i\rangle$ which has the Hartree shift subtracted off.
The term in wavy brackets can be viewed as a Green's-function weighted variance
of the bandstructure for $A_{\rm 1g}$ scattering, which can be interpreted
as a measure of the width of the strongly-correlated band.
Further, we note that in DMFT, where the self energy is local 
(and $d\rightarrow\infty)$, the 
sum rule for $B_{\rm 1g}$ Raman scattering would
also apply to the second moment of the optical conductivity:
\begin{equation}
\int_{0}^{\infty}d\nu \nu {\rm Im} \chi_{B_{1g}}({\bf q}=0,\nu)\propto
\int_{0}^{\infty}d\nu \nu^{2}\sigma(\nu).
\end{equation}
However, this relationship is violated if there is any momentum dependence associated with the self-energy or in finite dimensions.

\begin{figure}
\vspace{5mm}
\centerline{\psfig{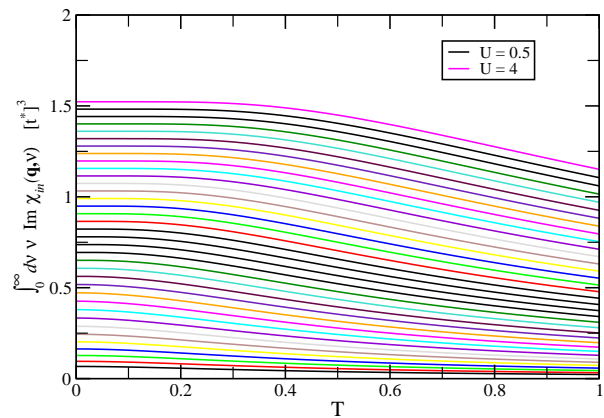}}
\vspace{.5cm}
\caption[]{$B_{\rm 1g}$ sum rules as a function
of temperature for the spinless FK model solved within DMFT
(the $A_{\rm 1g}$ results are similar).  The
different curves correspond to different
values of the interaction strength increasing in steps of 0.1 from 0.5
on the lowest curve to 4.0 on the highest curve.
}
\label{fig: fig1}
\end{figure}

We plot the sum rule for the FK model in Fig.~\ref{fig: fig1}  for $B_{\rm 1g}$
Raman scattering (${\bf q}=0$) and different values of $U$ as a function of 
temperature (results for $A_{\rm 1g}$ are similar). We have checked that
an explicit integral of the nonresonant Raman response agrees with the
sum rule plotted in Fig.~\ref{fig: fig1} for many different values of $U$
and $T$.
We expect similar results to hold for other models of correlated electrons 
like the Hubbard model. Fig.~\ref{fig: fig1} shows that these sum rules
are essentially constant at low temperature, indicating 
(i) the sum rule can be used to calibrate data from different samples
and different temperatures; (ii) the sum rule can be used to determine the     
frequency above which interband transitions become prominent; (iii)
the Raman response function multiplied by the frequency should be used to track
spectral weight shifts because of the sum rule; and (iv) the sum rules
have a momentum dependence for inelastic x-ray scattering that can
be calculated and compared to experiment.
Note that unlike the f-sum
rule, which is evenly weighted throughout the spectral range, the
Raman sum rules are heavily weighted at higher energies due to
the multiplication by the frequency.  
Consequently, unless there is a clear separation between low and high energy 
bands in the system, higher energy bands can significantly distort the sum rule.

The momentum dependence of the sum rule is reduced as $U$ increases and the
physical properties become more local.
The sum rule generally increases for
increasing momentum transfer since the kinetic energy term in 
Eq.~(\ref{eq: final_fk_result}) contributes to the
sum rule as phase space is created for light scattering by increasing ${\bf q}$.

\begin{figure}
\vspace{5mm}
\centerline{\psfig{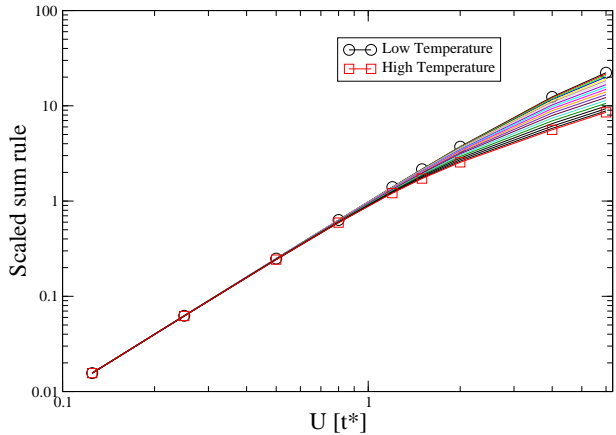}}
\vspace{.5cm}
\caption[]{$B_{\rm 1g}$ scaled sum rules showing the scaling behavior with
$U^{2}$ for the FK model discussed in the text. Deviations in scaling occur
when $U \ge U_{c}=\sqrt{2}t^*$, 
the critical value of $U$ for the metal-insulator transition
in the half-filled FK model.}
\label{fig: fig2}
\end{figure}

The underlying MIT is revealed via a simple scaling analysis, shown in
Fig.~\ref{fig: fig2}. It can be shown that
for metallic systems the sum rule varies as $U^{2}$ for small $U$, with large
deviations occuring as the MIT is approached. In Fig.~\ref{fig: fig2} we plot
the sum rule divided by $U^2$ for different temperatures as a function of $U$.
We see that the sum rule data collapses for small $U$ onto a single line for all
temperatures. The scaled data abruptly fans out from the straight line at approximately
$U=U_{c}$ and a strong temperature dependence emerges for larger $U$. Thus 
the onset of a strong temperature dependence and deviations from $U^{2}$ scaling
can be used as a straightforward and 
quantitative way to identify a MIT from the optical data.

\begin{figure}
\vspace{5mm}
\centerline{\psfig{file=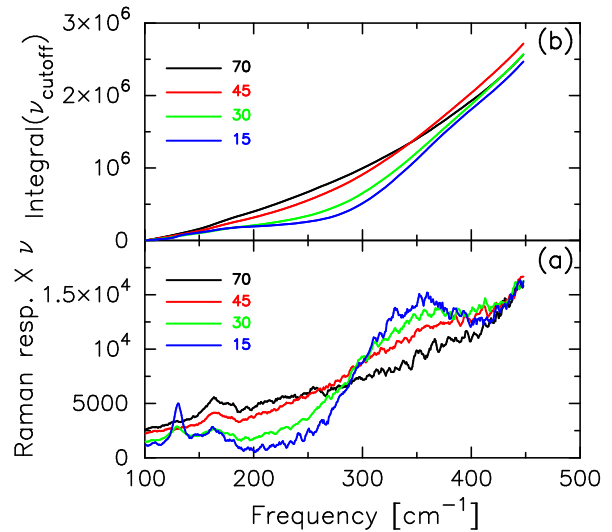,width=0.45\textwidth}}
\vspace{.5cm}
\caption[]{Raman response multiplied by frequency for SmB$_6$ at four
different temperatures.  On top is a plot of the integrated sum rule as
a function of the frequency cutoff. Both curves are plotted in arbitrary
units. The near crossing of the curves in 
the upper panel indicates where a cutoff can be placed (since the 
higher energy bands have little $T$ dependence, the
curves become parallel at high frequency, because the response is the
same for all $T$).
}
\label{fig: fig3}
\end{figure}

We show the relevance of these sum rules to an experimental
system that is ideal for this situation, SmB$_6$.  
The Raman results on SmB$_6$ provide an ideal comparison to this model for 
several reasons: the spectra have a clear separation between the low energy 
(intraband) and high energy (interband) scattering contributions; only the 
low energy component in SmB$_6$ exhibits significant spectral weight changes 
due to the gap formation at low temperatures; and the photon energy used in 
the experiment lies in a gap in the density of states of SmB$_6$, and 
therefore the Raman response is not likely to be influenced by resonant or 
mixed-scattering effects.  When we take the Raman data, and
multiply by the frequency, we find that the sum rule holds to
within five percent in this system (see Fig.~\ref{fig: fig3}), confirming the 
behavior determined from the FK model in DMFT (Fig.~\ref{fig: fig1}).

In summary, we have discovered a class of sum rules for inelastic light
scattering that are useful in analysing both Raman scattering and inelastic 
x-ray scattering.  The sum rules depend crucially on the form of the
electron interactions, and thereby yield useful information about the
correlations in the material.  
While light scattering data on correlated metals or insulators is still rather limited
(particularly with regard to inelastic X-ray scattering), our class of 
sum rules may be employed to analyze data and elucidate 
how the kinetic and potential energy of the 
system evolves as a function of doping, pressure, or 
temperature.

\end{document}